\documentclass[12pt,onecolumn,superscriptaddress,showpacs,preprintnumbers,amsmath,amssymb,prl,aps]{revtex4}

\usepackage{graphicx}
\usepackage{dcolumn}
\usepackage{bm}
\usepackage{color}

\newcommand{\TeV}{\ensuremath{\mathrm{Te\kern -0.1em V}}}
\newcommand{\TeVc}{\ensuremath{\mathrm{Te\kern -0.1em V\!/}c}}
\newcommand{\TeVcc}{\ensuremath{\mathrm{Te\kern -0.1em V\!/}c^2}}
\newcommand{\GeV}{\ensuremath{\mathrm{Ge\kern -0.1em V}}}
\newcommand{\GeVc}{\ensuremath{\mathrm{Ge\kern -0.1em V\!/}c}}
\newcommand{\GeVcc}{\ensuremath{\mathrm{Ge\kern -0.1em V\!/}c^2}}
\newcommand{\MeV}{\ensuremath{\mathrm{Me\kern -0.1em V}}}
\newcommand{\MeVc}{\ensuremath{\mathrm{Me\kern -0.1em V\!/}c}}
\newcommand{\MeVcc}{\ensuremath{\mathrm{Me\kern -0.1em V\!/}c^2}}

\newcommand{\myto}{\kern -0.3em\to\kern -0.2em}

\newcommand{\cdfii}{CDF\,II~}

\newcommand{\MM}{\mu^{+}\mu^{-}}
\newcommand{\JP}{B^{0}_{s} \rightarrow J/\psi \phi}
\newcommand{\PP}{B^{0}_{s} \rightarrow \psi(2S)\phi}
\newcommand{\JK}{B^{+} \rightarrow J/\psi K^{+}}
\newcommand{\PK}{B^{+} \rightarrow \psi(2S) K^{+}}
\newcommand{\JKS}{B^{0} \rightarrow J/\psi K^{*0}}
\newcommand{\PKS}{B^{0} \rightarrow \psi(2S) K^{*0}}
\newcommand{\JKL}{B^{0} \rightarrow J/\psi K^{0}}
\newcommand{\PKL}{B^{0} \rightarrow \psi(2S) K^{0}}
\newcommand{\jp}{J/\psi \phi}
\newcommand{\pp}{\psi(2S)\phi}

\newcommand{\Jmm}{J/\psi \rightarrow \mu^{+}\mu^{-}}
\newcommand{\Pmm}{\psi(2S) \rightarrow \mu^{+}\mu^{-}}
\newcommand{\Pmmpp}{\psi(2S) \rightarrow J/\psi \pi^{+}\pi^{-}}

\newcommand{\Bd}{B^{0}}
\newcommand{\Bu}{B^{+}}
\newcommand{\Bs}{B^{0}_{s}}

\begin{document}

\begin{flushright}
\end{flushright}

\title{Observation of $\PP$ and Measurement of Ratio of Branching Fractions  
${\cal B}(\PP)/{\cal B}(\JP)$ 
}

\affiliation{Institute of Physics, Academia Sinica, Taipei, Taiwan 11529, Republic of China} 
\affiliation{Argonne National Laboratory, Argonne, Illinois 60439} 
\affiliation{Institut de Fisica d'Altes Energies, Universitat Autonoma de Barcelona, E-08193, Bellaterra (Barcelona), Spain} 
\affiliation{Baylor University, Waco, Texas  76798} 
\affiliation{Istituto Nazionale di Fisica Nucleare, University of Bologna, I-40127 Bologna, Italy} 
\affiliation{Brandeis University, Waltham, Massachusetts 02254} 
\affiliation{University of California, Davis, Davis, California  95616} 
\affiliation{University of California, Los Angeles, Los Angeles, California  90024} 
\affiliation{University of California, San Diego, La Jolla, California  92093} 
\affiliation{University of California, Santa Barbara, Santa Barbara, California 93106} 
\affiliation{Instituto de Fisica de Cantabria, CSIC-University of Cantabria, 39005 Santander, Spain} 
\affiliation{Carnegie Mellon University, Pittsburgh, PA  15213} 
\affiliation{Enrico Fermi Institute, University of Chicago, Chicago, Illinois 60637} 
\affiliation{Joint Institute for Nuclear Research, RU-141980 Dubna, Russia} 
\affiliation{Duke University, Durham, North Carolina  27708} 
\affiliation{Fermi National Accelerator Laboratory, Batavia, Illinois 60510} 
\affiliation{University of Florida, Gainesville, Florida  32611} 
\affiliation{Laboratori Nazionali di Frascati, Istituto Nazionale di Fisica Nucleare, I-00044 Frascati, Italy} 
\affiliation{University of Geneva, CH-1211 Geneva 4, Switzerland} 
\affiliation{Glasgow University, Glasgow G12 8QQ, United Kingdom} 
\affiliation{Harvard University, Cambridge, Massachusetts 02138} 
\affiliation{Division of High Energy Physics, Department of Physics, University of Helsinki and Helsinki Institute of Physics, FIN-00014, Helsinki, Finland} 
\affiliation{University of Illinois, Urbana, Illinois 61801} 
\affiliation{The Johns Hopkins University, Baltimore, Maryland 21218} 
\affiliation{Institut f\"{u}r Experimentelle Kernphysik, Universit\"{a}t Karlsruhe, 76128 Karlsruhe, Germany} 
\affiliation{High Energy Accelerator Research Organization (KEK), Tsukuba, Ibaraki 305, Japan} 
\affiliation{Center for High Energy Physics: Kyungpook National University, Taegu 702-701; Seoul National University, Seoul 151-742; and SungKyunKwan University, Suwon 440-746; Korea} 
\affiliation{Ernest Orlando Lawrence Berkeley National Laboratory, Berkeley, California 94720} 
\affiliation{University of Liverpool, Liverpool L69 7ZE, United Kingdom} 
\affiliation{University College London, London WC1E 6BT, United Kingdom} 
\affiliation{Centro de Investigaciones Energeticas Medioambientales y Tecnologicas, E-28040 Madrid, Spain} 
\affiliation{Massachusetts Institute of Technology, Cambridge, Massachusetts  02139} 
\affiliation{Institute of Particle Physics: McGill University, Montr\'{e}al, Canada H3A~2T8; and University of Toronto, Toronto, Canada M5S~1A7} 
\affiliation{University of Michigan, Ann Arbor, Michigan 48109} 
\affiliation{Michigan State University, East Lansing, Michigan  48824} 
\affiliation{Institution for Theoretical and Experimental Physics, ITEP, Moscow 117259, Russia} 
\affiliation{University of New Mexico, Albuquerque, New Mexico 87131} 
\affiliation{Northwestern University, Evanston, Illinois  60208} 
\affiliation{The Ohio State University, Columbus, Ohio  43210} 
\affiliation{Okayama University, Okayama 700-8530, Japan} 
\affiliation{Osaka City University, Osaka 588, Japan} 
\affiliation{University of Oxford, Oxford OX1 3RH, United Kingdom} 
\affiliation{University of Padova, Istituto Nazionale di Fisica Nucleare, Sezione di Padova-Trento, I-35131 Padova, Italy} 
\affiliation{LPNHE-Universite Pierre et Marie Curie-Paris 6, UMR7585, Paris F-75005 France; IN2P3-CNRS} 
\affiliation{University of Pennsylvania, Philadelphia, Pennsylvania 19104} 
\affiliation{Istituto Nazionale di Fisica Nucleare Pisa, Universities of Pisa, Siena and Scuola Normale Superiore, I-56127 Pisa, Italy} 
\affiliation{University of Pittsburgh, Pittsburgh, Pennsylvania 15260} 
\affiliation{Purdue University, West Lafayette, Indiana 47907} 
\affiliation{University of Rochester, Rochester, New York 14627} 
\affiliation{The Rockefeller University, New York, New York 10021} 
\affiliation{Istituto Nazionale di Fisica Nucleare, Sezione di Roma 1, University of Rome ``La Sapienza," I-00185 Roma, Italy} 
\affiliation{Rutgers University, Piscataway, New Jersey 08855} 
\affiliation{Texas A\&M University, College Station, Texas 77843} 
\affiliation{Istituto Nazionale di Fisica Nucleare, University of Trieste/\ Udine, Italy} 
\affiliation{University of Tsukuba, Tsukuba, Ibaraki 305, Japan} 
\affiliation{Tufts University, Medford, Massachusetts 02155} 
\affiliation{Waseda University, Tokyo 169, Japan} 
\affiliation{Wayne State University, Detroit, Michigan  48201} 
\affiliation{University of Wisconsin, Madison, Wisconsin 53706} 
\affiliation{Yale University, New Haven, Connecticut 06520} 
\author{A.~Abulencia}
\affiliation{University of Illinois, Urbana, Illinois 61801}
\author{D.~Acosta}
\affiliation{University of Florida, Gainesville, Florida  32611}
\author{J.~Adelman}
\affiliation{Enrico Fermi Institute, University of Chicago, Chicago, Illinois 60637}
\author{T.~Affolder}
\affiliation{University of California, Santa Barbara, Santa Barbara, California 93106}
\author{T.~Akimoto}
\affiliation{University of Tsukuba, Tsukuba, Ibaraki 305, Japan}
\author{M.G.~Albrow}
\affiliation{Fermi National Accelerator Laboratory, Batavia, Illinois 60510}
\author{D.~Ambrose}
\affiliation{Fermi National Accelerator Laboratory, Batavia, Illinois 60510}
\author{S.~Amerio}
\affiliation{University of Padova, Istituto Nazionale di Fisica Nucleare, Sezione di Padova-Trento, I-35131 Padova, Italy}
\author{D.~Amidei}
\affiliation{University of Michigan, Ann Arbor, Michigan 48109}
\author{A.~Anastassov}
\affiliation{Rutgers University, Piscataway, New Jersey 08855}
% ...
\collaboration{CDF Collaboration}
\noaffiliation

\begin{abstract}
We report the first observation of $\PP$ decay in $p\overline{p}$ collisions at $\sqrt{s} = 1.96\,\TeV$ 
using  $360\,{\rm pb}^{-1}$ of data collected by the CDF II detector at the Fermilab Tevatron. 
We observe $ 20.2 \pm 5.0$ and $12.3 \pm 4.1$ $\PP$ candidates, in 
$\Pmm$ and $\Pmmpp$ decay modes, respectively. We present a measurement of the relative branching fraction 
${\cal B}(\PP)/{\cal B}(\JP)$ = 
$0.52 \pm 0.13({\rm stat.}) \pm 0.04({\rm syst.}) \pm 0.06({\rm BR})$ using the $\Pmm$ decay mode. 
\end{abstract}


\maketitle
\begin{figure}[ht]
\includegraphics[scale=0.4]{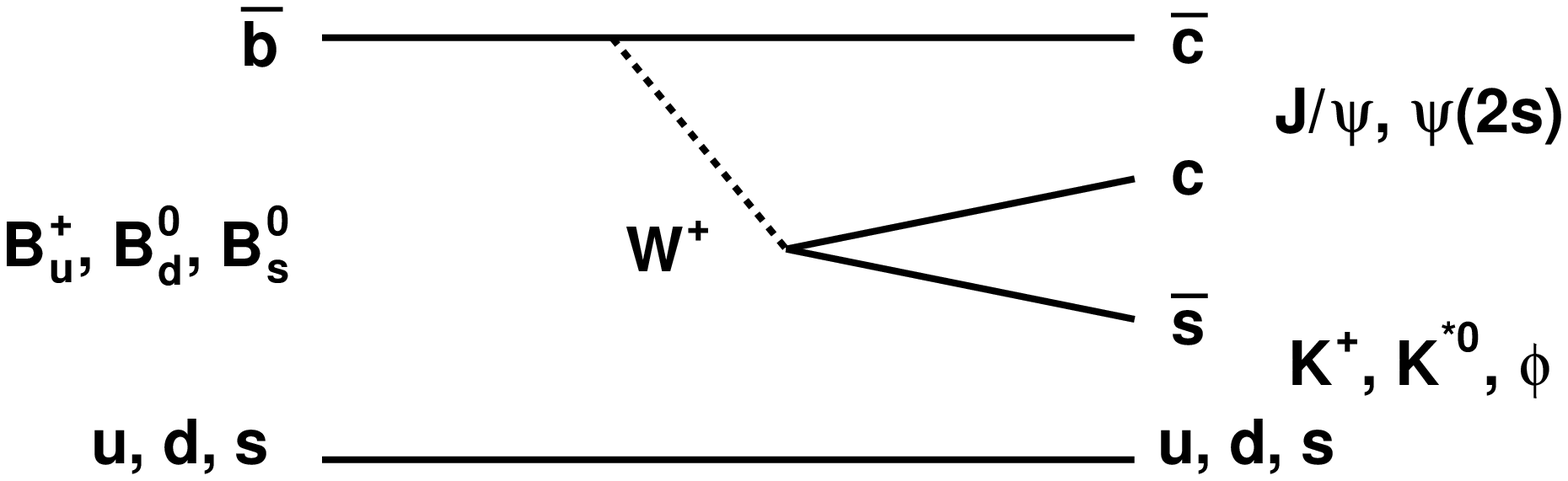}
\caption{Tree level Feynman diagram of $B$ mesons decaying to charmonium final states.}
\label{fig_FD}
\end{figure}

The decays of $B$ mesons to charmonium final states have been studied extensively in the past, 
and the measurements~\cite{babar, cleo, had} show that the ratio of the branching fractions of $\Bu$ and $\Bd$ 
decay to the $\psi(2S)$ final states over the $J/\psi$ final states are approximately 60$\%$ as shown 
in Table~\ref{current}.
The $B^{+, 0} \rightarrow J/\psi K^{+, *0}$  $(\psi(2S) K^{+, *0})$ and 
$\JP$ $(\psi(2S) \phi)$ are color-suppressed Cabibbo-favored decays 
that have the same tree-level decay topology as shown in Fig~\ref{fig_FD}. 
The relative branching ratio between $\PP$ and $\JP$ has not been measured. 
Only one $\PP$ candidate event has been reported at LEP in 1993~\cite{ALEPH93}.

\begin{table}[ht]
\caption{\label{current}The current relative branching ratio of $B$ meson decays between $\psi(2S)$ and $J/\psi$ final states.\\}
\begin{ruledtabular}
\begin{tabular}{lcc}
   Decay channel        & Value                         & Reference            \\
   \hline
   \Large $\frac{{\cal B}(\PK)} {{\cal B}(\JK)}$
                        & 0.64 $\pm$ 0.06 $\pm$ 0.07    & BaBar ~\cite{babar}  \vspace*{0.2cm} \\
  
   \Large $\frac{{\cal B}(\PKS) } {{\cal B}(\JKS)}$
                        & 0.61 $\pm$ 0.10               & PDG ~\cite{PDG2004}  \\
   \Large $\frac{{\cal B}(\PKL) } {{\cal B}(\JKL)}$
                        & 0.82 $\pm$ 0.13 $\pm$ 0.12    & PDG ~\cite{PDG2004}  \\
\end{tabular}
\end{ruledtabular}
\end{table}
\noindent

The $\JP$ mode has recently been used to determine the decay 
widths for the heavy and light $\Bs$ mass eigenstates by measuring the relative contribution of the CP-odd 
and CP-even components to the observed angular distribution as a function of the decay time~\cite{delta,d0}. 
Observing the $\PP$ would allow an independent measurement of 
the decay widths for the heavy and light $\Bs$ mass eigenstates in the future. 
In particular, the spin alignment of $\PP$ could be different from that of $\JP$.

In this Letter, we report the observation of $\PP$ in both $\Pmm$ and $\Pmmpp$ decay modes 
produced in $p\overline{p}$ collisions at $\sqrt{s} = 1.96\,\TeV$.
We also measure the ratio of branching fractions for $\JP$ and $\PP$.
Many systematic effects cancel in the measurement of the ratio, including uncertainties in 
total integrated luminosity, bottom-quark production and fragmentation, and trigger and 
reconstruction efficiencies. In addition, for this ratio of branching fractions measurement we use only the 
$\Pmm$ decay mode in order to guarantee identical topologies for
the $J/\psi$ and $\psi(2S)$ channels.
Similar decay modes, such as $\JK$ and $\PK$, are used as control samples to perform 
consistency checks, and study the systematic uncertainties.
Charge conjugate modes are implied throughout this paper. 
The data sample is comprised of about $3 \times 10^6$ $\Jmm$, $1 \times 10^5$ $\Pmm$, and 
$1.6 \times 10^4$ $\Pmmpp$ candidates.
The total integrated luminosity is approximately $360\,{\rm pb}^{-1}$ and was collected 
using the Collider Detector at Fermilab (CDF II) between February 2002 and July 2004.

The \cdfii detector is described in detail elsewhere~\cite{cdf}. 
The main components for this analysis are tracking and muon systems.
The tracks are reconstructed by the silicon microstrip detector (SVX\,II)~\cite{svxii}, 
the intermediate silicon layer (ISL)~\cite{isl}, 
and the Central Outer Tracker (COT)~\cite{cot}, which are immersed in a uniform axial 1.4 T magnetic field 
provided by a superconducting solenoid.
The SVX II is located at a radial distance between 2.5 and 10.6 cm, with double-sided micro-strip sensors 
arranged in five concentric cylindrical shells. The ISL provides one additional layer for central tracks, and two for tracks with $|\eta| > $ 1.0.
The layers are located radially between the SVX II and the COT.
The COT is a multi-wire drift chamber, 
which consists of 96 layers of sensitive wires, grouped into 8 superlayers. These superlayers cover 
the radial region between 40 cm and 137 cm from the beam, and are arranged to provide alternating axial and 
$\pm 2^{ \circ}$ tilted stereo measurements.
Planar drift chambers~\cite{cmu} located outside the calorimeter
are used to identify muons in the central region ($|\eta| < $ 1.0, where $\eta$ is the pseudorapidity). 
The events are selected with a three-level trigger system. 
At Level~1, charged particle trajectories in the plane transverse to the beam direction are reconstructed from the COT hits using a hardware processor~\cite{xft}.
The trigger requires tracks with transverse momentum ${p_T}(\mu) > 1.5$ GeV/$c$ to be matched to hits in the muon detector.
At Level~2, opening angle and opposite-charge cuts are imposed on the muon pairs.
At Level~3, the two muon tracks are required to be oppositely charged with invariant mass between 
2.7 and 4.0 GeV/$c^{2}$. 


We reconstruct $\JP$ and $\PP$ followed by $\Pmm$ and $\Pmmpp$, where $\Jmm$ and $\phi \to K^+K^-$.
For the measurement of the relative branching fraction between $\JP$ and $\PP$, 
it is desirable to have selection criteria similar for both decay modes. 
All three $\Bs$ decay channels involve only the well known $J/\psi$, $\psi(2S)$ and $\phi$ decays, 
which have been used extensively in other measurements at CDF, and their selection criteria are well established. 
In this analysis, we follow the selection requirements developed in 
the $b$ hadron mass measurements~\cite{other1} and apply them to the three $\Bs$ decay modes of interest. 

The reconstruction begins by selecting $\Jmm$ or $\Pmm$ candidates, 
with pairs of oppositely charged tracks that satisfy the muon pair trigger requirements. 
The reconstructed $\MM$ invariant mass is required to be within 
80 MeV/$c^2$ of the $J/\psi$ or $\psi(2S)$ mass~\cite{PDG2004}. 
The $\Pmmpp$ is reconstructed by associating a $\Jmm$ candidate (with its mass constrained to the $J/\psi$ mass) with a pair of 
tracks, each with $p_T > $ 0.4 GeV/$c$.  
The invariant mass of $J/\psi \pi^+ \pi^-$ is required to be within 20 MeV/$c^2$ of the world average 
$\psi(2S)$ mass~\cite{PDG2004}. 
Once a $J/\psi$ or $\psi(2S)$ candidate is selected, 
we search  for a $\phi \to K^+K^-$ candidate with a pair of additional tracks. The invariant mass of $K^{+}K^{-}$ 
is required to be within 10 MeV/$c^2$ of the $\phi$ mass~\cite{PDG2004}. 
The $p_T$ of the $\phi$  candidate is required to be greater than 2.0 GeV/$c$.  
The $\Bs$ meson candidates are then reconstructed by associating a $J/\psi$ or $\psi(2S)$ candidate with 
a $\phi$ candidate. 
All tracks (4 tracks in $\JP$ or $\PP$ followed by $\Pmm$ and 6 tracks in $\PP$ followed by $\Pmmpp$) 
are required to be consistent with having originated from a common vertex satisfying vertex quality requirements.
Prompt background, with tracks coming directly from the primary vertex, can be reduced by exploiting variables
sensitive to the long lifetime of the $\Bs$ meson. To reduce prompt background, 
the transverse decay length ($L_{xy}$) of the $\Bs$ is required to exceed 100 $\mu m$, 
where $L_{xy}$ is defined as the transverse vector from the beam axis to the $\Bs$ decay vertex projected 
onto the transverse momentum of the $\Bs$ candidate. 
To ensure a well measured $B$ meson decay vertex, 
each track is required to have a measurement in at least three axial layers of the silicon detector, including SVXII and ISL.
The transverse momentum of the $\Bs$ candidates is required to be greater than 6.5 GeV/$c$ to further 
reduce combinatoric background.
To improve the $B$ meson mass resolution, the $\mu^{+}\mu^{-}$  mass is constrained to the $J/\psi$ or 
$\psi(2S)$ mass, while the $J/\psi \pi^{+}\pi^{-}$ mass is constrained to the $\psi(2S)$ mass.

\begin{figure*}[ht]
\centering
\includegraphics[scale=0.3]{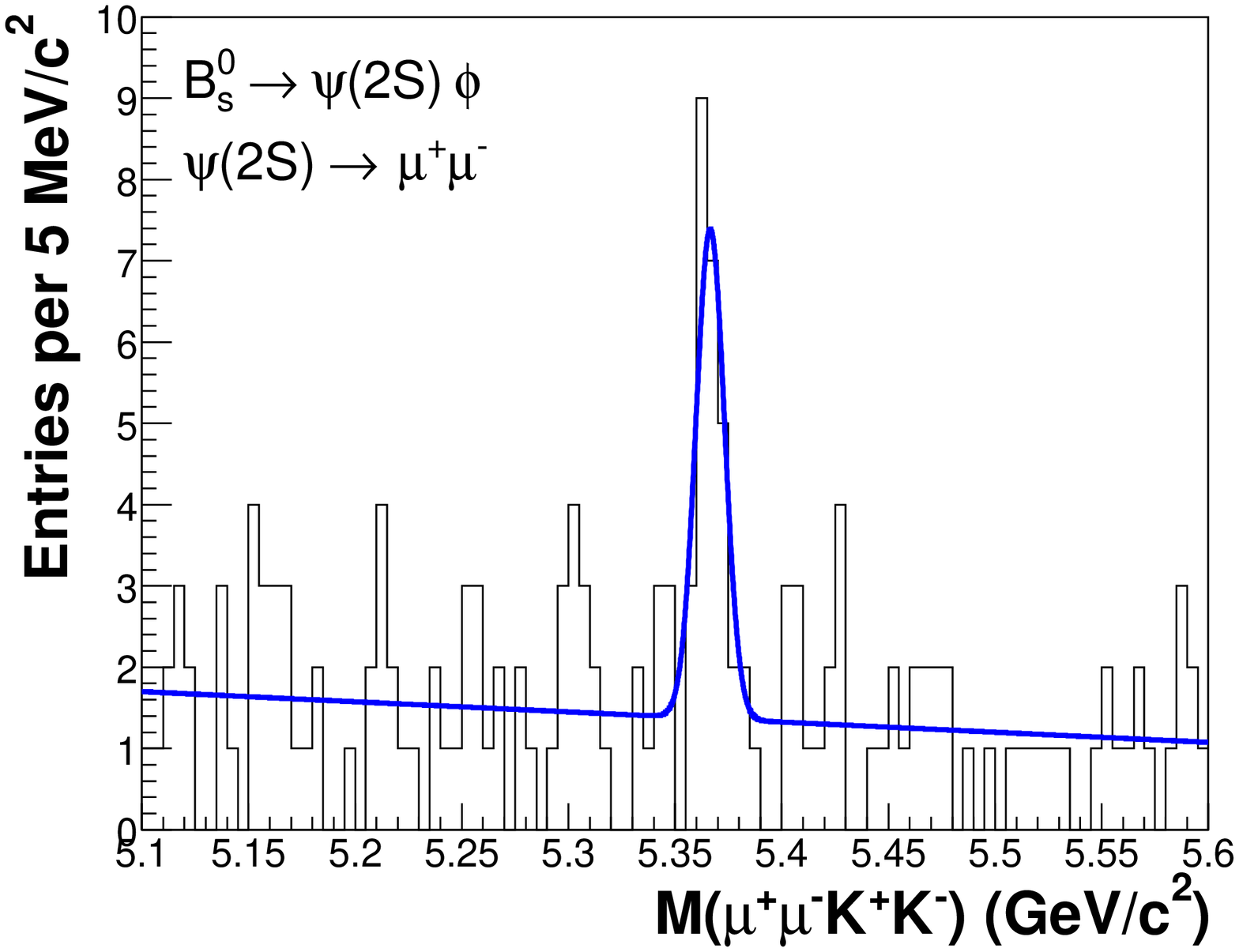}
\includegraphics[scale=0.3]{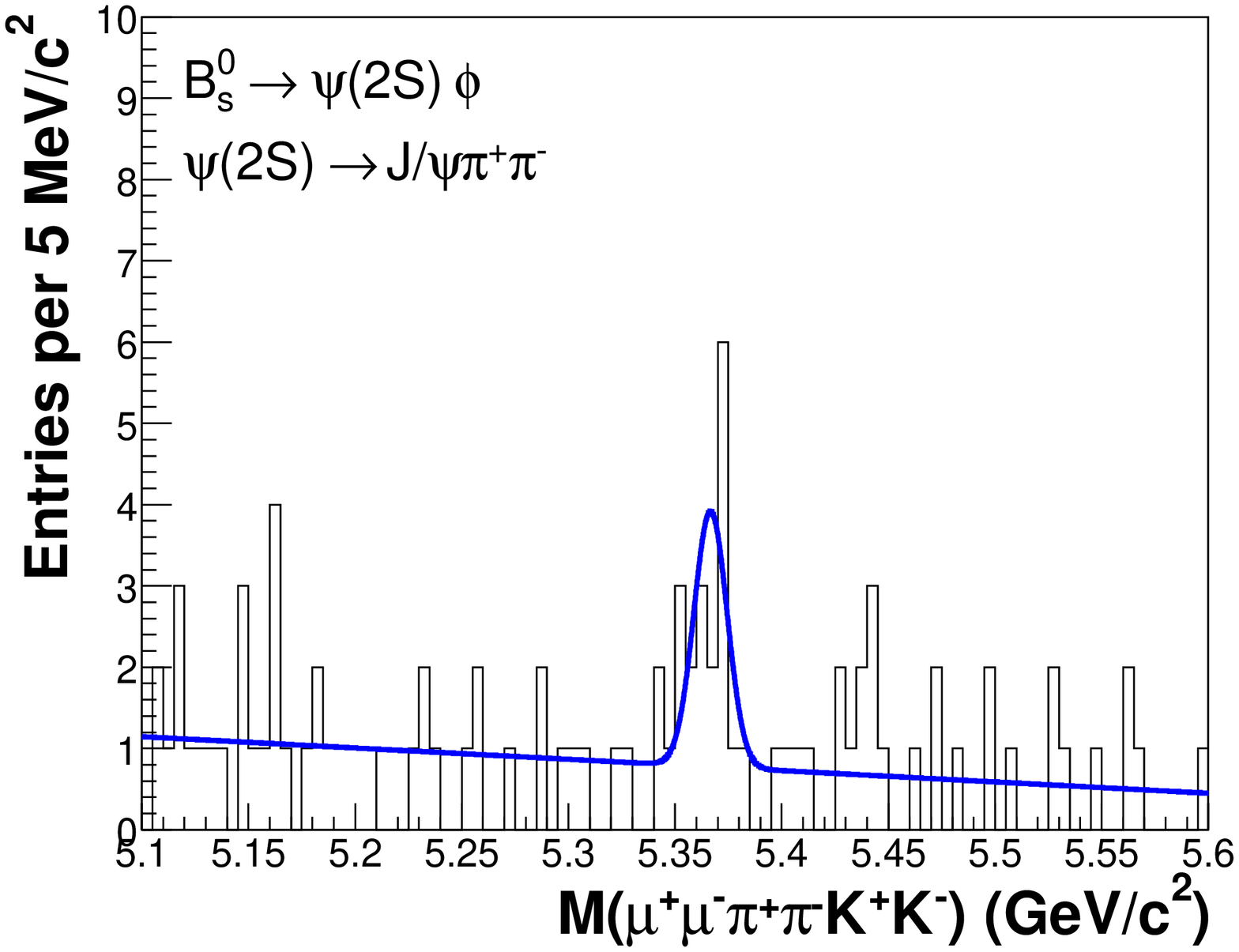}
\includegraphics[scale=0.3]{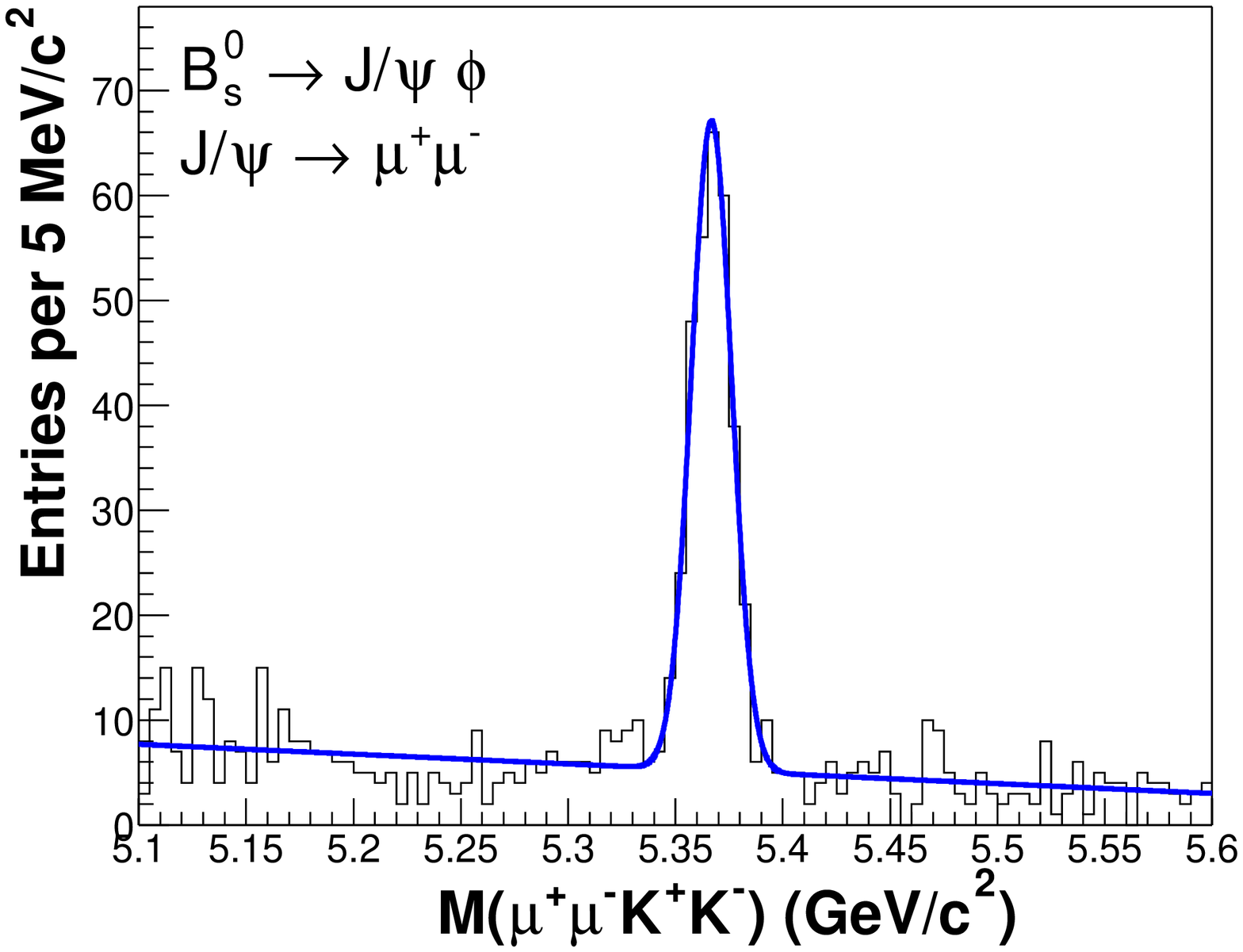}
\caption{\label{fig_PP}
Invariant mass distributions for $\JP$ (bottom), and for $\PP$, followed by $\Pmm$ (upper left), 
or $\Pmmpp$ (upper right).
The curves are the results of the fits described in the text.}
\end{figure*}

Two sources of background are expected in the $\Bs$ signal region: combinatoric background and 
kinematic ``reflection'' of $\JKS$ (for $\JP$) or $\PKS$ (for $\PP$), where the pion from the 
$K^{*0}$ decay is mis-assigned as a kaon. The combinatoric background is modeled by a first order polynomial. 
The $\JKS$ ($\PKS$) reflection background results in a broad distribution near and above the $\Bs$ signal region. 
The fraction of $\JKS$ ($\PKS$) events that fall into the $\JP$ ($\PP$) signal region is estimated 
using a Monte Carlo simulation. 
The background contribution from reflection in our data sample is then calculated 
by multiplying the fraction determined from Monte Carlo simulation by the number of the $\JKS$ and
$\PKS$ candidates in the same data. The contribution of the $\JKS$ reflection in the $\JP$ signal region 
is estimated to be $6.6 \pm 0.3$ events. The contribution of the $\PKS$ reflection in the $\PP$ signal region 
is estimated to be $0.34 \pm 0.05$ and $0.19 \pm 0.03$ events for $\Pmm$ and $\Pmmpp$ modes, respectively.
The $\JKS$ ($\PKS$) reflection background is highly suppressed because only a small fraction of the misidentified 
$K^{*0} \to K^+\pi^-$ can satisfy the $\phi \to K^+K^-$ mass requirement.

\begin{table*}[ht]
  \caption{The numbers of observed signal events, the fitted masses, and the signal width (Gaussian sigma, or mass resolution) for each of the 
three $\Bs$ decay channels. 
Note that the width for each of $\PP$ decay modes is fixed as described in the text.\\
 \label{yields}}
\begin{ruledtabular}

\begin{tabular}{lccc}
	Decay          &  Mean[MeV/$c^2$]  & Width[MeV/$c^2$] 	& Yield \\
   \hline
         $\JP$         & 5366.76$\pm$0.66  &  9.42$\pm$0.58  		&  292.2$\pm$ 15.9   \\
         $\PP;\Pmm$    & 5366.50$\pm$1.86  &  6.63 (fixed)   		&  20.2 $\pm$5.0  \\  
         $\PP;\Pmmpp$  & 5366.63$\pm$3.20  &  7.77 (fixed)   		&  12.3$\pm$4.1 \\ 
\end{tabular}
\end{ruledtabular}
\end{table*}

An unbinned log-likelihood fit is used to extract signal yields from the reconstructed mass spectra, 
as shown in Fig.~\ref{fig_PP}. The signal distribution is modeled as a Gaussian, 
and the background distribution is modeled as a first order polynomial. 
The background component from misidentified  $K^{*0}$ decays
is also included, with a shape obtained from the Monte Carlo simulation.
The width for each of the two $\PP$ modes is fixed in the following way. We take the ratio of the widths 
for $\PP$ relative to $\JP$ as a scale factor determined from Monte Carlo simulation, 
and we then calculate the width for $\PP$, using the width of $\JP$ from data. 
Comparison between Monte Carlo and data for the control samples of $\JK$ and $\PK$ shows that the relative ratio 
of the widths of the two modes can be well predicted by Monte Carlo simulation.
The signal yields, fitted masses, and width (Gaussian sigma, or mass resolution) of the three decay channels are summarized 
in Table~\ref{yields}. A consistency check (Monte Carlo independent) is performed by fitting the $\phi \to K^+K^-$ invariant mass spectra 
for events in the $B_{s}$ signal region after sideband subtraction. The $\phi \to K^+K^-$ signal yield obtained this way in data is consistent 
with that from fitting the $B_{s}$ mass spectra, indicating that the $B^0$ reflection background from $\JKS$ ($\PKS$) or other 
decay modes is indeed negligible..

The background contribution in the signal region 
(defined as a window six times the expected mass resolution, as shown in Table II, 
around the mean value of the $B_{s}$ signal peak) for $\PP$, followed by $\Pmm$ and $\Pmmpp$ decays,  
is estimated to be 10.0 $\pm$ 3.2 and 6.5 $\pm$ 2.6 events, respectively. 
The probability of a statistical fluctuation of the expected total background in the signal region to the 
observed or higher number of events is 2.5 $\times$ $10^{-7}$ for $\PP$ with $\Pmm$ and 1.6 $\times$ $ 10^{-5}$ 
for $\Pmmpp$. 
These correspond to 5.0 $\sigma$ and 4.2 $\sigma$ one-sided Gaussian significance for the two decay modes, 
respectively.
The combined probability of the two modes is 1.1 $\times$ $ 10^{-10}$, corresponding to a 6.4 $\sigma$ 
significance for the observation of $\PP$.


We measure the relative branching fraction between $\JP$ and $\PP$ using only the $\Jmm$ decay mode, and the control sample
data ($\JK$ and $\PK$) are used to study the systematic uncertainties.
The relative branching ratio for the $\MM$ mode is extracted using the formula: 
\begin{widetext}
\begin{equation}
\begin{split}
\label{eq:Brbs}
\frac{{\cal B}(\PP)}{{\cal B}(\JP)} = 
\frac{N_{\pp}}{N_{\jp}} \cdot \frac{{\cal B}(\Jmm)}{{\cal B}(\Pmm)} \cdot \frac{\epsilon_{\jp}}{\epsilon_{\pp}},
\end{split}
\end{equation}
\end{widetext}
\noindent where $\epsilon_{\jp}/\epsilon_{\pp} = 0.925 \pm 0.006$ is the ratio of the combined trigger and selection
efficiencies derived from Monte Carlo simulation (with the error due to the size of the simulated samples), 
and $N_{\jp}$ or $N_{\pp}$ is the total number of 
reconstructed $\Bs$ mesons for each mode. The ${\cal B}(\Jmm)$ and ${\cal B}(\Pmm)$ 
are the world average branching fractions~\cite{PDG2004}.

In our analysis, we use a Monte Carlo simulation to determine the relative efficiency 
for the two decay modes, and the control sample data are used to study the systematic uncertainties. 
The simulation of the CDF II detector is based upon a {\sc geant} description~\cite{geant}. 
Transverse momentum and rapidity distributions of single $b$ quarks are 
generated based on next-to-leading-order (NLO) pertubative QCD~\cite{bquark}. 
The $\Bs$ meson spectrum used in the Monte Carlo simulation is consistent with the data from inclusive $B \to J/\psi X$~\cite{cdf}. 
The {\sc evtgen} program~\cite{evtgen} is used to decay $B$ mesons into the final states of interest.


Since both modes are $\Bs$ decays, and the decay topologies are very similar,  
most systematic uncertainties cancel in the ratio. Systematic uncertainties originate from 
fitting the invariant mass distributions to obtain signal yields, from determination of the relative efficiencies,
 and from the measured branching fractions of $J/\psi$ and $\psi(2S)$ decays taken from~\cite{PDG2004}. 
Consistency checks are performed on the fitting method by varying the range and
using different functions, and no statistically significant variation is found.
Systematic uncertainty from the fitting method is evaluated by performing the fitting without constraints 
on the width.

Systematic uncertainties on the ratio of efficiencies are due to the differences in the 
kinematics of the two decay modes. For example, due to the mass difference 
between $\psi(2S)$ and $J/\psi$, the $p_T$($\mu$) distributions are somewhat different between
the two decay modes. To take into account the difference in $p_T$($\mu$) distributions, 
the single muon efficiency measured from data~\cite{meffi} is used to reweight the 
Monte Carlo samples, and the relative efficiency (central value and error) is recalculated. 
We vary the measured muon efficiency, and find that the systematic uncertainty due to the difference 
in $p_T$($\mu$) distributions is negligible. 
The main systematic uncertainty due to decay kinematics difference comes from lack of 
knowledge of the angular correlation in the $\PP$ decay.
The central value of the relative efficiency is determined by assuming that the angular 
correlation of the $\PP$ decay is the same as that of the $\JP$.
To evaluate the effects on our measurement, we generate 
Monte Carlo samples with pure CP-even and CP-odd decays  
for $\PP$ and recalculate the relative efficiency. We take the difference between
CP-even and CP-odd cases as the systematic uncertainty, which turns out to be the major component ($5.5\%$).
The systematic uncertainty from the fitting contributes at the $3.9\%$ level. 
The total systematic uncertainty is $6.7\%$.

The contribution from the branching fractions is calculated by propagating the world average uncertainties.
The dominant contribution is due to the measured 
branching ratio ${\cal B}(\psi(2S) \rightarrow \mu^{+}\mu^{-}) = (0.73 \pm 0.08)\%$~\cite{PDG2004}.


Using Eq.~\ref{eq:Brbs}, we derive the ratio of relative branching fractions:
\begin{widetext}
\begin{equation}
\label{eq:result}
\frac{{\cal B}(\PP)}{{\cal B}(\JP)} = 
0.52 \pm 0.13({stat.}) \pm 0.04({syst.}) \pm 0.06({BR})
\end{equation}
\end{widetext}
\noindent where the first error is statistical, the second is systematic, and the third is due to 
the branching ratios of $\Jmm$ and $\Pmm$.

In summary, we present the first observation of $\PP$ decay, in both $\Pmm$ and $\Pmmpp$ modes in $p\overline{p}$ collisions at 
$\sqrt{s} = 1.96\,\TeV$ using the \cdfii detector. We also present the measurement of the ratio of branching fractions 
between $\PP$ and $\JP$ using the $\Pmm$ decay mode. This result for $\Bs$ is consistent with the ratios of 
branching fractions for the corresponding decays of $\Bu$ and $\Bd$~\cite{PDG2004}, indicating that the relative branching 
ratio of $B$ meson decays between $\psi(2S)$ and $J/\psi$ final states is independent of the flavor of the lighter quark.

We thank the Fermilab staff and the technical staffs of the participating institutions for their vital contributions. This work was supported by the U.S. Department of Energy and National Science Foundation; the Italian Istituto Nazionale di Fisica Nucleare; the Ministry of Education, Culture, Sports, Science and Technology of Japan; the Natural Sciences and Engineering Research Council of Canada; the National Science Council of the Republic of China; the Swiss National Science Foundation; the A.P. Sloan Foundation; the Bundesministerium f\"ur Bildung und Forschung, Germany; the Korean Science and Engineering Foundation and the Korean Research Foundation; the Particle Physics and Astronomy Research Council and the Royal Society, UK; the Russian Foundation for Basic Research; the Comisi\'on Interministerial de Ciencia y Tecnolog\'{\i}a, Spain; in part by the European Community's Human Potential Programme under contract HPRN-CT-2002-00292; and the Academy of Finland.


\end{document}